\documentclass[%
 reprint,
 amsmath,amssymb,
 aps,
floatfix,
]{revtex4-1}

\usepackage{graphicx}
\usepackage{dcolumn}
\usepackage{bm}
\usepackage{hyperref}
\usepackage{color}
  
\usepackage{ulem}
\usepackage{color}

\begin{document}

\title{First-principles identification of single-photon emitters based on carbon clusters in hexagonal boron nitride}.

\author{C. Jara}
\affiliation{LAAS-CNRS, Université de Toulouse, CNRS, Toulouse, France}

\author{T. Rauch}
\affiliation{Institut f\"{u}r Festk\"{o}rpertheorie und -optik, Friedrich-Schiller-Universit\"{a}t Jena, Max-Wien-Platz 1, 07743 Jena, Germany}

\author{Silvana Botti}
\affiliation{Institut f\"{u}r Festk\"{o}rpertheorie und -optik, Friedrich-Schiller-Universit\"{a}t Jena, Max-Wien-Platz 1, 07743 Jena, Germany}
\affiliation{European Theoretical Spectroscopy Facility}

\author{Miguel A. L. Marques}
\affiliation{Institut für Physik, Martin-Luther-Universität Halle-Wittenberg, 06120 Halle/Saale, Germany}
\affiliation{European Theoretical Spectroscopy Facility}

\author{A. Norambuena}
\affiliation{Centro de Investigaci\'on DAiTA Lab, Facultad de Estudios Interdisciplinarios, Universidad Mayor, Chile}

\author{R. Coto}
\affiliation{Centro de Investigaci\'on DAiTA Lab, Facultad de Estudios Interdisciplinarios, Universidad Mayor, Chile}

\author{J. R. Maze}
\affiliation{Institute of Physics, Pontificia Universidad Católica de Chile, Santiago, Chile}
\affiliation{Research Center for Nanoscale and Advanced Materials (CIEN), Pontificia Universidad Católica de Chile}

\author{F. Munoz}
\email{fvmunoz@gmail.com}
\affiliation{Center for the Development of Nanoscience and  Nanotechnology, CEDENNA, Santiago, Chile}
\affiliation{Departamento de F\'isica, Facultad de Ciencias, Universidad de Chile, Santiago Chile}

\date{\today}

\begin{abstract}
A recent study associate carbon with single photon emitters (SPEs) in hexagonal boron nitride (h-BN). This observation, together with the high mobility of carbon in h-BN suggest the existence of SPEs based on carbon clusters. Here, by means of density-functional theory calculations we studied clusters of substitutional carbon atoms up to tetramers in hexagonal boron nitride. Two different conformations of neutral carbon trimers have zero-point line energies and shifts of the phonon sideband compatible with typical photoluminescence spectra. Moreover, some conformations of two small C clusters next to each other result in photoluminescence spectra similar to those found in experiments. We also showed that vacancies are unable to reproduce the typical features of the phonon sideband observed in most measurements due to the large spectral weight of low-energy breathing modes, ubiquitous in such defects.
\end{abstract}

\maketitle

\section{Introduction}

Single photon emitters (SPEs) in large bandgap materials are emerging as promising building blocks for quantum information processing and distribution. The interplay between their optical and spin properties has enabled several high-impact quantum technology applications\cite{Smith2019} such as quantum sensing \cite{Degen2017,acsnano1}, and quantum computation \cite{Rong2015,acsnano2}. Therefore, a growing community is on an active search for long coherence spin systems based on bright and stable single emitters. However, although many new emitters have been characterized in diamond \cite{Sipahigil2016,Ekimov2015,Palyanov2015,Iwasaki2017}, silicon-carbide\cite{Castelletto2020} and hexagonal boron nitride (h-BN) \cite{Castelletto2020b}, unrevealing their atomic structure in order to take advantage of their properties is lagging behind. In particular, a number of new emitters have been found in h-BN \cite{tran16,Wigger19,kozawa2019,Martinez16,Du2015,Mendelson2020,Jung17,acsnano3,acsnano4} with different brightness, stability and zero phonon lines and phonon sidebands. In this work we calculate the expected emission spectrum for several defects in h-BN and associate some of these defects with the observed photoluminescence found in recent works, specifically we are interested in the SPEs with a zero-phonon line (ZPL) in the near-infrared or visible spectrum (1.6 to 2.4 eV) \cite{Wigger19}.

Despite extensive theoretical attempts to identify the atomic structure associated with SPE defects found in h-BN~\cite{tran16nnat,tawfik19,sajid2019,atta11,kozawa2019,ivady2019,sajid2020,abdi18}, up to now few defect configurations have been proposed. The theoretical work has mostly relied on density functional theory (DFT) under the assumption that it can provide transition energies accurate enough to estimate the zero-phonon line (ZPL) and compare it to experimentally measured values~\cite{toth19}. The \textit{de facto} standard for DFT calculations of defects in h-BN is the Heyd–Scuseria–Ernzerhof hybrid functional (HSE06)~\cite{hse03,hse06} since it is one of the exchange-correlations functional giving the most accurate estimation for the band gap of  solids~\cite{pedro19}. It predicts a band gap of 5.7 eV in h-BN, close to the experimental value of 6.1 eV~\cite{Cassabois2016,Elias2019}. However, it is worth noting that the average error of HSE06 in determining insulator band gaps is about 30\% (or 0.7 eV)~\cite{pedro19}. Therefore, HSE06 (and DFT in general) is not accurate enough to predict signatures of electronic transitions in defects. This issue was addressed for a single defect in h-BN by comparing HSE06 with more accurate \textit{ab initio} methods~\cite{Reimers18}. Errors range from 0 to 1 eV depending on the orbitals involved (closed or open shell, triplet or singlet states, etc). 

In view of this relatively large estimation error, instead of identifying defects through their ZPL, we focus on other features such as their phonon sidebands and other experimental/theoretical considerations on their formation energy. In particular, we focus on a commonly found relative shift of the phonon sideband (PSB) from the ZPL of $\sim$ 160 meV \cite{Du2015}, in order to determine the families of defects compatible with this characteristic feature and narrow down the potential PSE to few possibilities.
Moreover, a recent study \cite{Mendelson2020} shows that carbon impurities are related to the observed SPEs in h-BN. This observation, together with the high mobility of carbon impurities in a h-BN layer \cite{weston18}, suggest carbon clusters as an interesting alternative to explore for the identification of SPEs. 

This article is organized as follows. In Section \ref{sec:evidence} we summarize experimental evidence about the SPEs in h-BN including energetics of defects formation and migration. Our methodology is explained in Section \ref{sec:methods}, we check our methodology against the 4.1 eV SPE, identified as the C$_\mathrm{B}$C$_\mathrm{N}$ defect \cite{sinke19}. In Section~\ref{sec:vacancies} we elaborate on the observed PSB and the nature of the defects, ruling out defects with vacancies. In Section \ref{sec:clusters} we study C neutral and charged clusters up to tetramers, identifying neutral C trimers as candidates for the SPEs and providing insight into the electronic structure of the clusters. Finally, in Section~\ref{sec:doublecluster} we study defects composed of two nearby C clusters (double clusters), identifying five defects compatible with experimental data.

\section{Current Evidence of SPE in h-BN}
\label{sec:evidence}

Most photoluminescence (PL) experiments have shown several visible (or near-infrared) SPEs in h-BN, with ZPLs ranging from $\sim1.63$ to $\sim2.15$ eV~\cite{tran16,Wigger19,kozawa2019,Martinez16}. Regardless of the ZPL energy, most of the SPEs have a similar PSB, featuring a first maximum at about $160\pm 10$ meV from the ZPL~\cite{tran16,jung16}. Another study~\cite{Wigger19} found a larger spread of this maximum (150-200 meV). These energies are very large for a single phonon, and they are found only in the LO phonons of (pristine) h-BN~\cite{Wigger19}. Therefore, the atomic rearrangement during the emission process should have a strong bond stretching (\textit{i.e.} associated with a high-energy phonon). This is the most important ingredient of our model and we will elaborate on it in Sec.~\ref{sec:vacancies}.

Another feature, shared among the reported PL spectra of SPEs in h-BN, is a sharp but asymmetrical ZPL. This hints an important contribution of low-energy phonons (50 meV or less) in the PL \cite{Wigger19}.

No experimental work is conclusive about the nature of the luminescent defects, notwithstanding there exist several clues about them. Despite the similitude of the PL spectra, it was shown that at least 4 different defects are responsible for the spectra found in the 540- to 720 nm range \cite{hayee2020}. Furthermore, the SPEs exhibit a giant stark effect, that suggest to exclude defects with C$_3$ and inversion symmetries~\cite{xia2019,nikolay19,noh18}. Finally, Mendelson~\textit{et al.}~\cite{Mendelson2020} showed that the presence of C impurities is fundamental to have SPEs in h-BN. In addition, they found a much larger concentration of C--B rather than C--N bonds in samples with SPEs. This does not necessarily rule out defects with several C--N bonds as SPEs, but they are less likely to occur.

Other studies gives inconclusive evidence about the identification of SPEs. For the sake of completeness we comment on them. Feng {\it et al.}~\cite{feng18} found a large concentration of vacancies in an h-BN sample by optical imaging methods, but the only evidence relating vacancies to SPEs are DFT calculations of the ZPL. Kozawa {\it et al.}~\cite{kozawa2019} followed an experimental and theoretical approach, including STEM imaging, kinetic Monte-Carlo, and DFT calculations to postulate isomers of vacancies as the SPEs. They found no evidence of heavier heteroatoms in the samples. They explained the $\sim 160$ meV PSB with the bulk phonon of this specific energy~\cite{Cheja16}.

Finally, it is worth commenting on few studies about the relevant formation energies and energy barriers of defects in h-BN. According to Weston \textit{et al.} \cite{weston18}, vacancies in h-BN have very large formation energy. In a B-rich environment, the N vacancy V$_\mathrm{N}$ has a formation energy above 2.5 eV (depending on the position of the Fermi level). The B vacancy V$_\mathrm{B}$ has an even higher formation energy of more than 5.5 eV in the most favorable environment. The combination of a vacancy and an antisite, N$_\mathrm{B}$-V$_\mathrm{N}$, has a formation energy larger than 8 eV. In the case of few-layer h-BN, inter-layer intrinsic defects are possible \cite{Strand2019}, the N inter-layer defect has a defect formation energy close to 2 eV with a Fermi level close to the CBM. The formation energy of the B inter-layer defect is very large regardless of the Fermi level. The formation energy of defects involving C atoms in h-BN is energetically much more favorable \cite{weston18}. In a N-poor (N-rich) environment the formation energy of the substitutional defect C$_\mathrm{N}$ (C$_\mathrm{B}$) varies from $-\sim -1.5$- to $\sim 1.5$ eV, depending on the Fermi level. Therefore, C impurities are likely to be present. The energy of an interstitial carbon C$_i$ is large, ranging from 2.5 to 7.0 eV, depending on the position of the Fermi level. But the migration barrier of C$_i$ is just 1.0 eV, therefore it can roam almost freely until its coalescence with another defect, probably with a vacancy \cite{weston18}.

\section{Methodology}
\label{sec:methods}
First-principles calculations were performed with DFT using the Vienna ab initio simulation package (VASP)~\cite{vasp1,vasp2,vasp3,vasp4} with projector augmented-wave pseudopotentials~\cite{paw}. The code pyProcar was used for post-processing the DFT output \cite{HERATH2020}. We used a kinetic energy cutoff of 400 eV. Our supercell is made of 98 atoms, which corresponds to $7\times 7$  unit cells of the single layer of h-BN. We used a single k-point ($\Gamma$) in the supercell calculations. These convergence parameters ensure an accuracy of 3 meV/atom on the total energy. In most calculations -including structural relaxations- we used the HSE06 exchange-correlation functional~\cite{hse06}, after pre-converging using a less processing demanding calculation made with the Perdew-Burke-Ernzerhof (PBE)~\cite{pbe} exchange-correlation functional. A force tolerance of 0.01 eV/\AA was used for optimizing the geometry in both ground and excited states. To identify allowed electronic transitions, we examined the imaginary part of the dielectric tensor. The procedure to calculate the excited levels consists of fixing the occupation numbers of the wave functions \cite{Gali09,gavnholt08}. This is often referred to as the $\Delta$SCF method in the literature. 

For the calculation of vibrational modes, we used the PBE exchange-correlation functional~\cite{pbe}, since it provides a reliable description of phonons at a lower computational cost~\cite{favot90,Alkauskas2014}. The size of the supercell is crucial to have a correct PL spectrum~\cite{Alkauskas2014}, therefore we checked the convergence of the spectral function with the supercell size.

In a seminal paper, Alkauskas \textit{et al.}~\cite{Alkauskas2014} determined the PL spectrum of a single N vacancy NV$^-$ center in diamond by using the spectral function derived from the electron-phonon coupling~\cite{miyakawa70}. We will briefly summarize the procedure to obtain the spectral function $S(\hbar \omega)$, following~\cite{Alkauskas2014}. For further details, please refer to that article. The difference in atomic positions between the (relaxed) excited and ground state is represented by a generalized set of coordinates $q_k$. Decomposing $q_k$ in the system's vibrational modes, $k$:

\begin{equation}
    q_k = \sum_{\alpha i} \sqrt{m_\alpha}\left(R_{e,\alpha i} - R_{g,\alpha i}\right)r_{k,\alpha i},
\end{equation}
\noindent where $k,\alpha$, and $i$ are labels for vibrational modes, atoms, and Cartesian coordinates, respectively. $m_\alpha$ is the mass of the atom $\alpha$. The (relaxed) positions of the ground and excited states are $R_{g}$ and $R_{e}$, respectively. The $k$-th vibrational mode is the unitary vector $r_k$. Here, we have assumed that the phonons of the ground and excited states are identical. From the $q_k$, it is possible to define a partial Huang-Rhys factor
\begin{equation}
    S_k = \frac{\omega_k q_k^2}{2\hbar},
\end{equation}
with $\omega_k$ the frequency of mode $k$. This leads to the spectral function
\begin{equation}
    S(\hbar\omega) = \sum_k S_k\delta(\hbar\omega-\hbar\omega_k),\label{eq:sk}
\end{equation}
with $\delta(x)$ the Dirac delta function, which for practical purposes is replaced by Gaussians with a width of 6 meV~\cite{Alkauskas2014}. Eq.~(\ref{eq:sk}) provides direct access to the PSB. The prominent peak of $S(\hbar\omega)$ indicates a strong coupling between the phonon $k$ with $q_k$, and hence with $q$. Therefore, a PBS peak at $160$ meV from the zero-phonon line (in the PL spectrum) translates in a peak of $S(\hbar\omega)$ at $160$ meV. To obtain the PL spectrum we need the time-dependent spectral function
\begin{equation}
    S(t) = \int_0^\infty S(\hbar\omega)e^{-i\omega t} \,d(\hbar\omega).
\end{equation}
The optical spectral function $A(\hbar\omega)$ is the Fourier transform of the generating function $G(t) = e^{S(t)-S(0)}$, \textit{i.e.}
\begin{equation}
    A(E_{ZPL} - \hbar\omega) = \frac{1}{2\pi}\int_{-\infty}^{\infty} G(t)e^{i\omega t - \gamma |t|} dt,\label{eq:pl}
\end{equation}
where the ZPL energy, $E_{ZPL}$, enters as a parameter. The other parameter is $\gamma$, which represents a thermal broadening of the ZPL and is chosen to fit the experimental PL spectrum. Nevertheless, we will show that regardless of the value of this parameter, it is not possible to reproduce the measured PL spectrum for some defects such as vacancies. Unless stated otherwise, we set $\gamma=4$ (1/ps). The luminescence intensity, $L(\omega)$, is $L(\hbar\omega)\propto \omega^3A(\hbar\omega)$. 

\subsection{The 4.1 \lowercase{e}V SPE}
\label{sec:test}

\begin{figure}[t]
    \centering
    \includegraphics[width=\columnwidth]{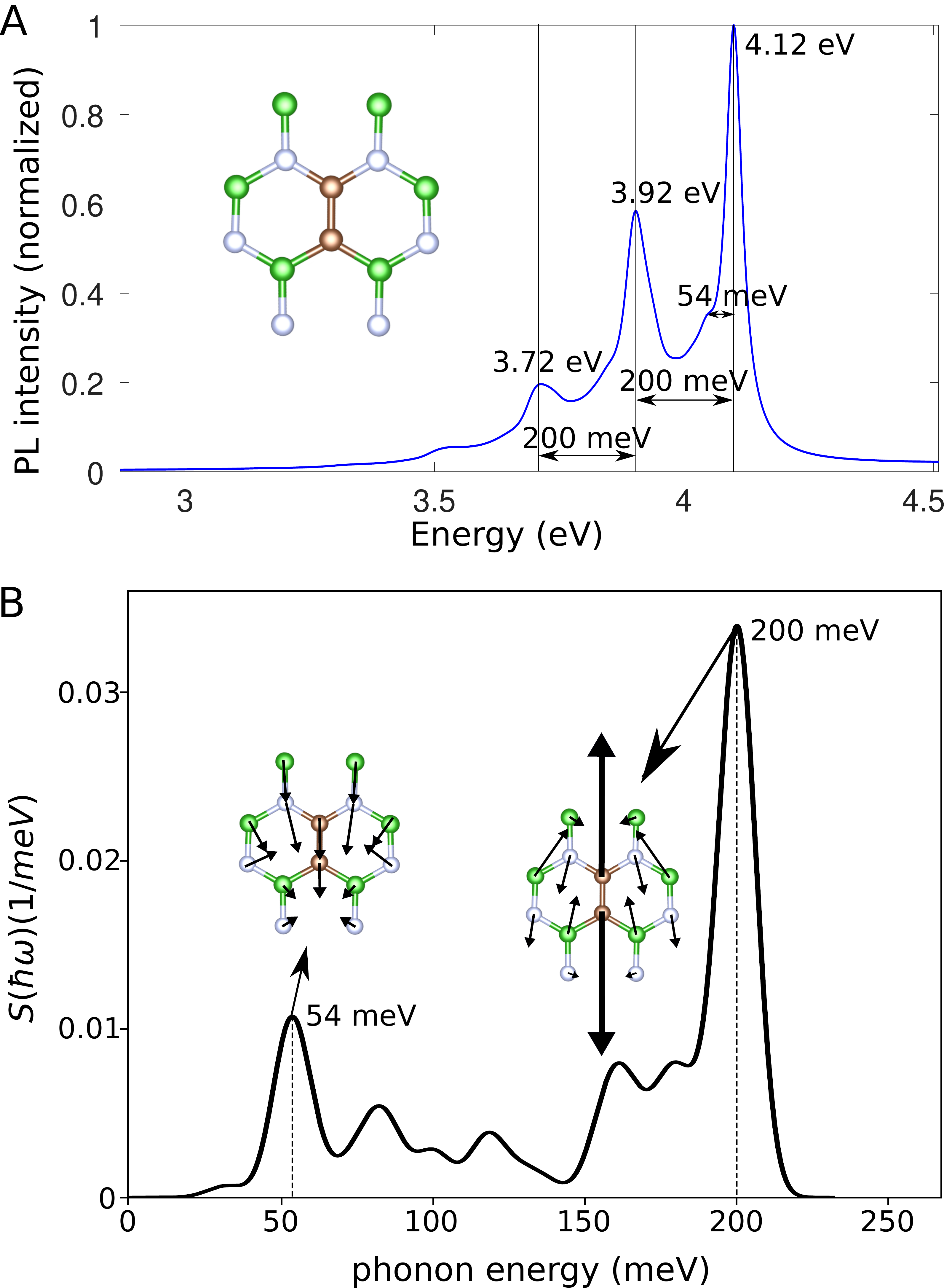}
    \caption{(a) Simulated PL spectrum (solid lines) of the C$_\mathrm{B}$C$_\mathrm{N}$, the vertical lines signal the energy (in eV) of the measured peaks \cite{Du2015}. The inset shows the geometry of the C$_\mathrm{B}$C$_\mathrm{N}$ defect (gray: C, green: B, light blue: N) (b) The spectral function of the defect C$_\mathrm{B}$C$_\mathrm{N}$, the inset near each maximum represents the respective phonon modes. The horizontal arrows indicate the energy of the most important phonons in $S(\omega)$, and the corresponding phonon shift in the PL spectrum.}
    \label{fig:c-dimer}
\end{figure}

Experimentally a deep ultraviolet SPE, at 4.12 eV, was measured~\cite{Du2015} in h-BN. It features peaks of the PSB at 3.92 eV and 3.72 eV. Theoretically, it was identified as the C$_\mathrm{B}$C$_\mathrm{N}$ defect \cite{sinke19}, based on the involvement of carbon atoms, the transition energies, and the Huang-Rhys factor. DFT calculations --with the HSE exchange-correlation \cite{hse03}-- give a ZPL of 4.31 eV \cite{sinke19}, close to the experimental value. 
Additionally, Korona and Chojecki~\cite{korona19} calculated the relevant phonon modes of this defect, finding a strong stretching with an energy of $\sim 200$ meV, in agreement with the experimental PL. Because of the evidence linking the C$_\mathrm{B}$C$_\mathrm{N}$ with the 4.12 emission line, we will use it as a benchmark of our methodology.

The defect geometry is shown in the inset of Fig.~\ref{fig:c-dimer}a. Using the methodology of Sec~\ref{sec:methods}, we calculated the spectral function and the PL spectrum of the C$_\mathrm{B}$C$_\mathrm{N}$ defect, see Fig.~\ref{fig:c-dimer}. The ZPL was set to the experimental value, \textit{i.e.} 4.12 eV, since our approach relies on finding the shape of the PBS rather than  accurately determining the ZPL. We found an excellent agreement between our model and the experimental PL. The peaks of the PSB at $\sim 200$ meV are easy to identify from the spectral function, $S(\hbar\omega)$, see Fig~\ref{fig:c-dimer}b. It features a prominent peak exactly at the energy of the first phononic replica of the PSE. This phonon corresponds to a bond stretching. In general, only bond stretchings have a large enough energy and, therefore, they are required to emulate the experimental PL. A second --smaller-- peak is just over the 50 meV and it does not involve any obvious bond stretching, see Fig.~\ref{fig:c-dimer}. These low-energy phonon modes leads to an asymmetric ZPL such as that found in \cite{Wigger19}.

\section{Vacancies}
\label{sec:vacancies}

\begin{figure}[h!]
    \centering
     \includegraphics[width=0.9\columnwidth]{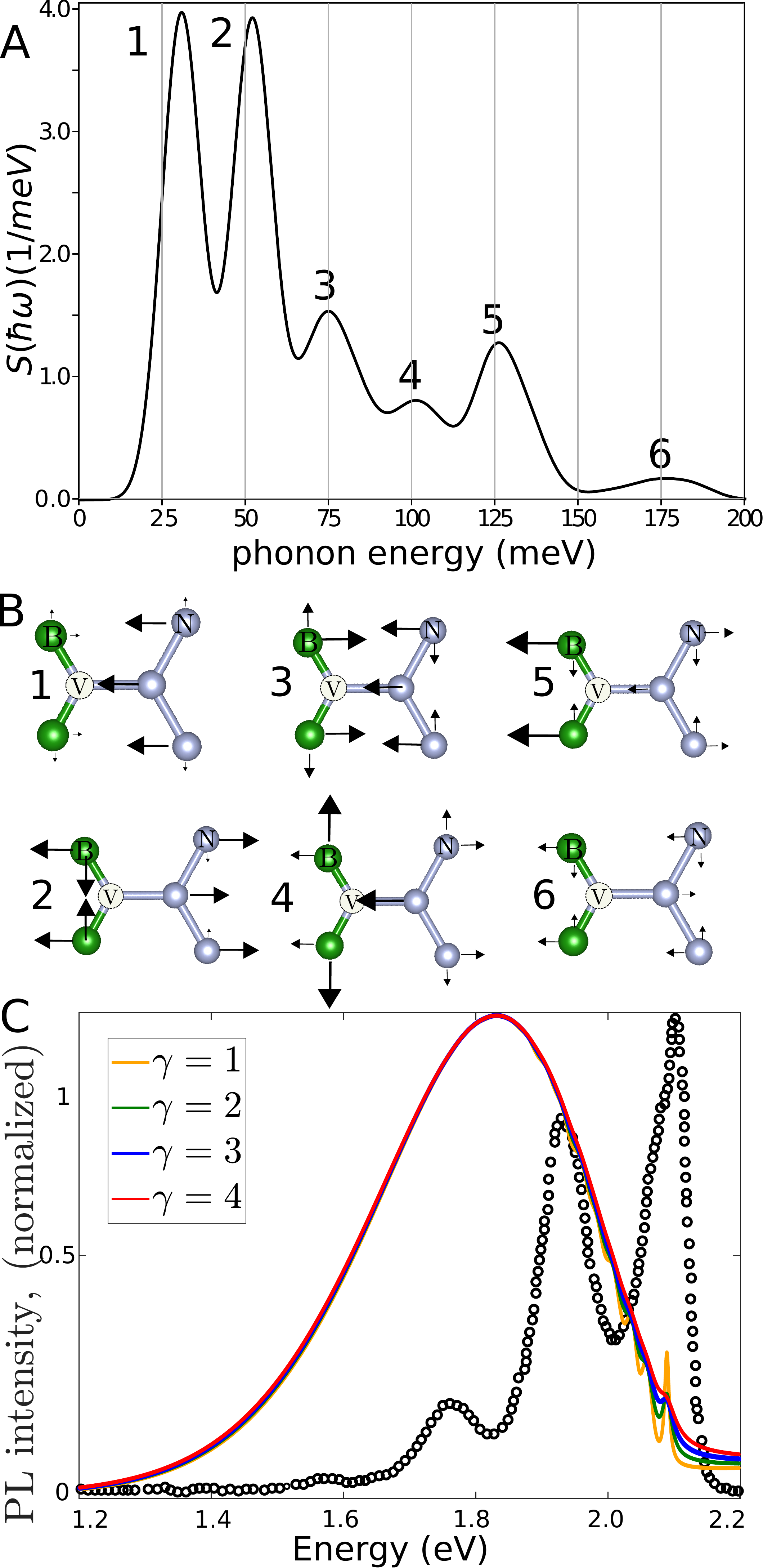}
    \caption{(a) Spectral function $S(\omega)$ of the V$_\mathrm{N}$N$_\mathrm{B}$. The numbers at each peak are the respective phonon modes of panel (b). (c) PL spectra of V$_\mathrm{N}$N$_\mathrm{B}$, obtained by Eq.~\ref{eq:pl}, for several values of the parameter $\gamma$, in (1/ps). A typical PL spectrum, from Ref.~\cite{Martinez16}, is added (black circles) to allow the comparison between the PL associated with a defect including vacancies and the experimental one. Regardless of the value of $\gamma$, it is not possible to fit the experimental PL spectrum by a V$_\mathrm{N}$N$_\mathrm{B}$.}
    \label{fig:vacancies}
\end{figure}

Several studies have focused on SPEs composed of a vacancy plus a substitutional or antisite atom, probably motivated by the color centers in diamond NV$^-$ and SiV. In this section, we argue that this kind of defect is incompatible with the observed PSB. 

The defect V$_\mathrm{N}$N$_\mathrm{B}$ was proposed as a possible SPE in h-BN because its ZPL has a good agreement with the experimental value \cite{tran16nnat}. The spectral function of this defect, $S(\omega)$,  see Fig.~\ref{fig:vacancies}a, features prominent peaks for low-energy phonons. These phonons correspond to the spectral projection of the atomic rearrangement of a transition between the ground and excited states. In this case -and in general, for defects involving vacancies- the most important phonon modes are the breathing modes, and they are low-energy, coordinated modes involving several atomic displacements without a clear bond stretching. A scheme of these modes is presented in Fig.~\ref{fig:vacancies}b. The PL spectrum, calculated from $S(\omega)$, has two free parameters (see Eq.~\ref{eq:pl}): the energy of the ZPL, which is adjusted to coincide with experiments, and $\gamma$, which sets the broadening of the modes. Regardless of the value of $\gamma$, it is impossible to exhibit the PSB shift of $\sim 160$ meV with this defect. For the sake of comparison, a typical PL spectrum of SPEs in h-BN (from Ref.\cite{Martinez16}) is added in Fig.~\ref{fig:vacancies}c (black circles). 

\begin{figure}
    \centering
    \includegraphics[width=0.95\columnwidth]{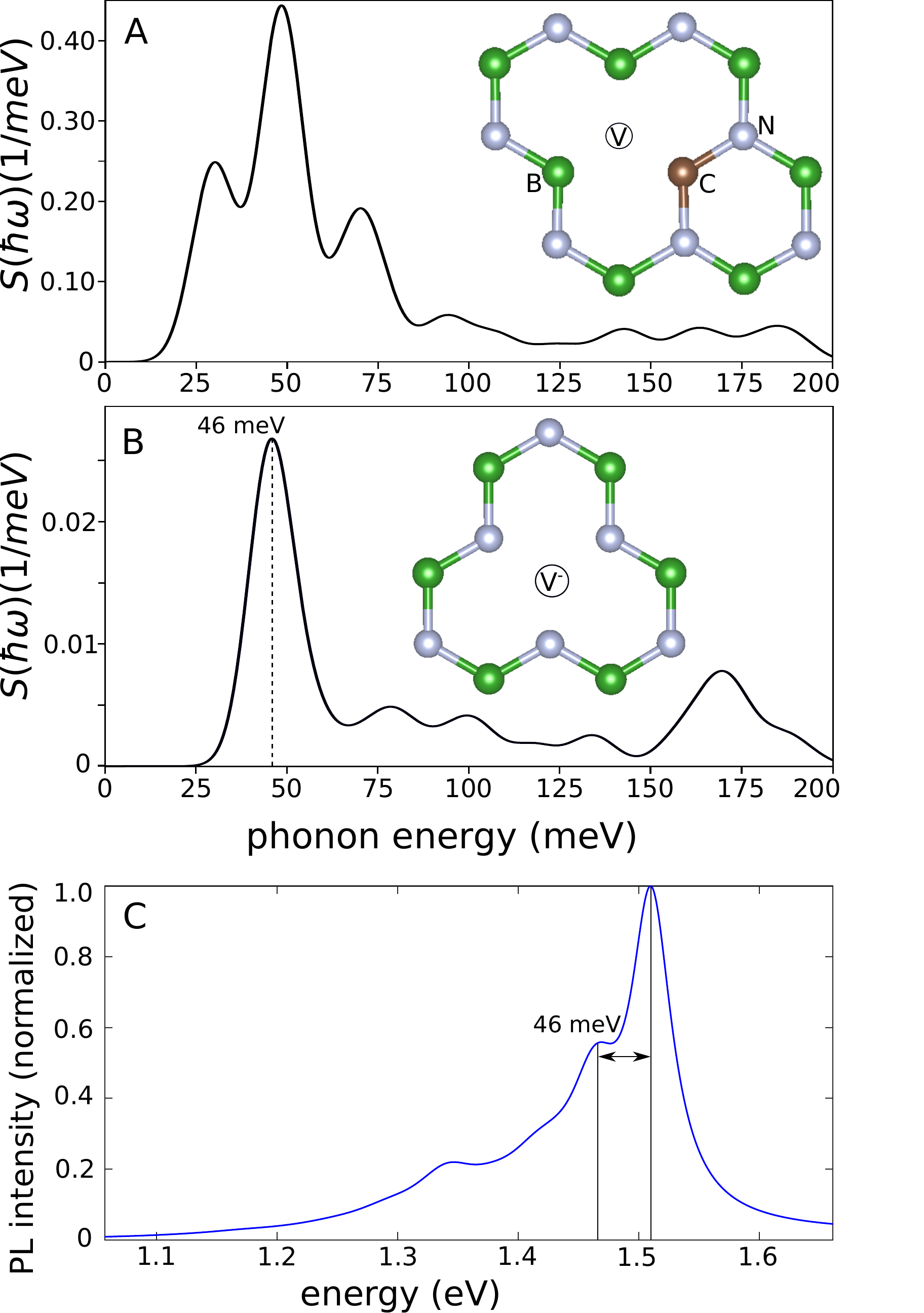}
    \caption{Spectral function $S(\omega)$ and defect configuration for the  V$_\mathrm{N}$C$_\mathrm{B}$ (panel A) and V$_\mathrm{B}^-$ (panel B). All the prominent peaks correspond to low-energy phonons. PL spectrum of the V$_\mathrm{B}^-$ defect. The most important phonon shift is indicated in panels B and C.}
    \label{fig:morevacancies}
\end{figure}

Similarly, the spectral function of the V$_\mathrm{N}$C$_\mathrm{B}$ defect has prominent low-energy modes, but it lacks of high-energy phonons needed to reproduce the PSB, see Fig.~\ref{fig:morevacancies}a. Therefore, we can rule out this defect as one of the SPEs we are interested in \cite{tran16,Wigger19,kozawa2019,Martinez16,Mendelson2020,Jung17}.

Recently, the negatively charged boron vacancy (V$B^-$) was atomically engineered \cite{kianinia2020}. Its PL spectrum shows a single broad maximum. Its PSB has a width of $\sim 137$ meV after the ZPL. In Figs.~\ref{fig:morevacancies}b and ~\ref{fig:morevacancies}c, $S(\omega)$ and the calculated PL spectra are shown, in qualitative agreement with the experimental data \cite{kianinia2020}. Once again, the single wide asymmetric peak is explained by the prevalence of low-energy phonons.

\section{C clusters}
\label{sec:clusters}

\subsection{Carbon monomers}

The simplest C impurities are single substitutional defects, C$_\mathrm{N}$ and C$_\mathrm{B}$. While they are not SPEs (their electron-phonon interaction and symmetry are incompatible with the experimental data, \textit{i.e.} not C$_{3}$ symmetry)
, they are helpful to understand more complex C clusters. The following discussion will be qualitative since we are interested in finding trends. Later in this section, when we find potential SPEs, we will provide quantitative information. 

\begin{figure}
    \centering
    \includegraphics[width=0.9\columnwidth]{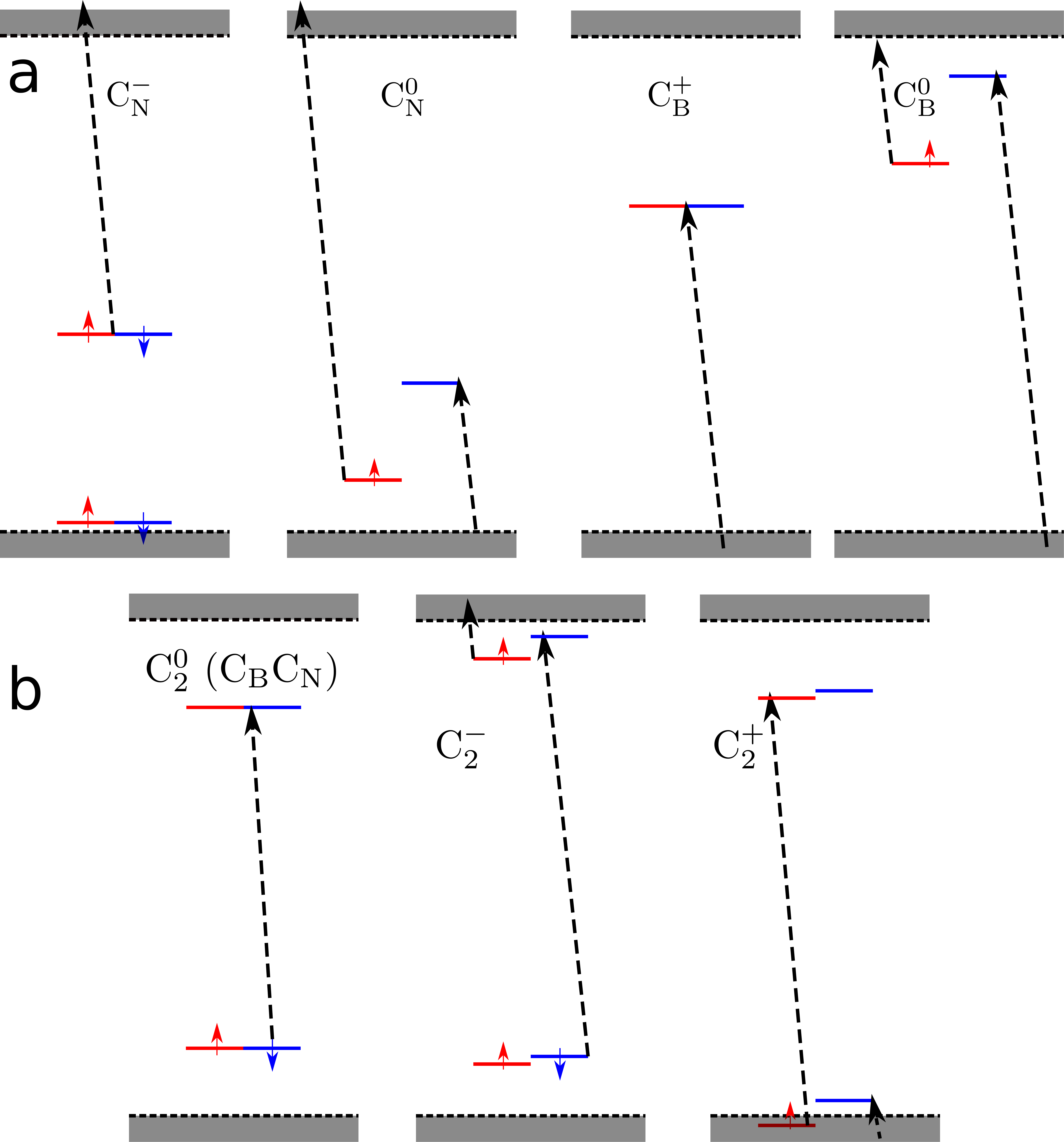}
    \caption{Schemes of the energy levels of the C$_\mathrm{N}$ and C$_\mathrm{B}$ defects (a) and the C$_2$ defect (b). The gray regions indicate the bulk valence and conduction bands. The shorter lines within the band gap are defect states. The color denotes the spin and occupied levels have an arrow. The dashed tilted arrows denote possible optical transitions. The extra defect level of C$_\mathrm{N}^-$ is buried in the bulk valence bands in C$_\mathrm{N}^0$.}
    \label{fig:monomers}
\end{figure}

In the case of C$_\mathrm{N}$, the C atom has one electron less than the N atom, therefore it creates an n-type (acceptor) energy level just above the valence band maximum. For C$_\mathrm{B}$ it is the converse, it creates a p-type (donor) energy level just below the conduction band minimum. The charged defects C$_\mathrm{N}^-$ and C$_\mathrm{B}^+$ are iso-electronic to pristine h-BN, therefore their defect energy levels (within the bulk band gap) are fully occupied and empty, respectively, see Fig.~\ref{fig:monomers}a. This implies that the optically allowed transitions have energies larger than the observed range in PL experiments. A neutral C$_\mathrm{N}^0$ or C$_\mathrm{B}^0$ defect has a half-filled defect band, and therefore, non-degenerate 
, see Fig.~\ref{fig:monomers}a. They have two sets of allowed transitions: 
from their occupied defect state to the conduction band, or from the valence band to an empty defect state.

\subsection{Carbon dimer}

In order to simplify the notation for the next discussion, we will denote the C$_\mathrm{B}$C$_\mathrm{N}$ dimer as C$_2$. The stable configuration of C$_2$ is neutral. It has two defect levels within the band gap: one is fully occupied while the other is empty, see Fig.~\ref{fig:monomers}b. This gives origin to the 4.1 eV SPE, see Sec.~\ref{sec:test}. The charged dimers C$_2^+$ and C$_2^-$ have a half-occupied defect state similar to the  neutral C monomers. However, unlike the monomers, the defect states are closer to the valence band maximum and conduction band minimum, making them unsuitable to be the SPEs in the range we are interested (\textit{i.e.}, the optical transitions have an energy lower than 1 eV for one spin, but larger than 3.5 eV for the other spin). Therefore, we can discard them as SPEs in the visible range.

\subsection{Carbon trimers}

\begin{figure}
    \centering
    \includegraphics[width=0.8\columnwidth]{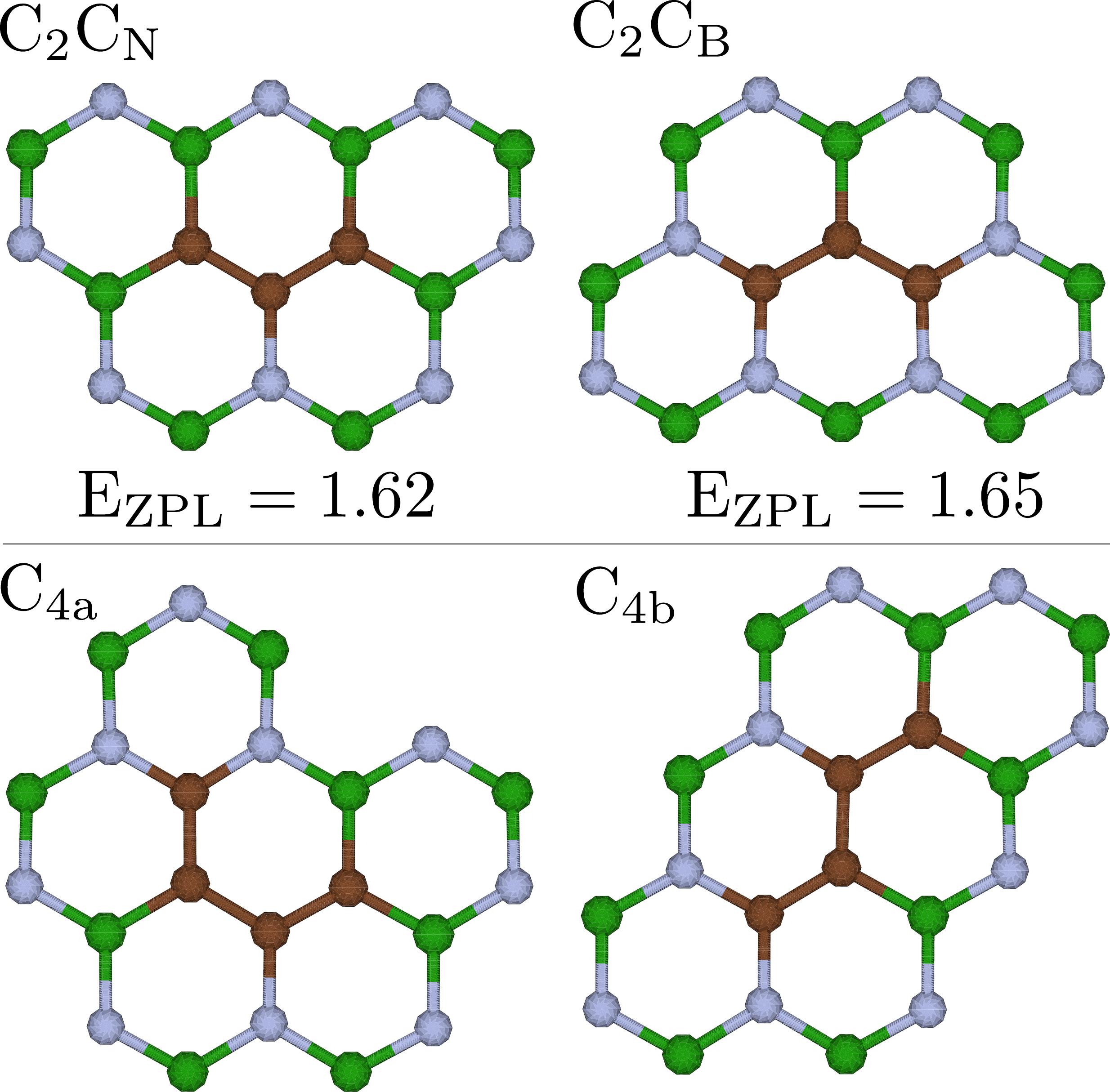}
    \caption{Trimers (upper row) and tetramers (lower row) of substitutional C defects. The energy $E_{ZLP}$ of the neutral trimers is given in eV, see the text for details.}
    \label{fig:tri-tetra}
\end{figure}

C trimers can be thought of as a monomer adjacent to a dimer so that their wavefunction are mixed sharing features common to both subsystems. Two conformations are possible, C$_2$C$_\mathrm{B}$ and C$_2$C$_\mathrm{N}$, with C$_2$ being C$_\mathrm{B}$C$_\mathrm{N}$. As in  monomers, the C$_2$C$_\mathrm{B}^+$ and C$_2$C$_\mathrm{B}^-$ are isoelectronic to pristine h-BN (\textit{i.e.} they have the same number of electrons of a h-BN layer), hence, their defect states within the fundamental band gap are fully occupied (C$_2$C$_\mathrm{N}^-$) or empty (C$_2$C$_\mathrm{B}^+$). This is similar to the schemes of the charged monomers in Fig.~\ref{fig:monomers}a. Unlike the monomers, there are more defect levels within (or close to) the band gap. The allowed optical transitions have an (vertical) energy of at least 3 eV, too large to be add the SPEs we are searching for. 

\begin{figure}
    \centering
    \includegraphics[width=\columnwidth]{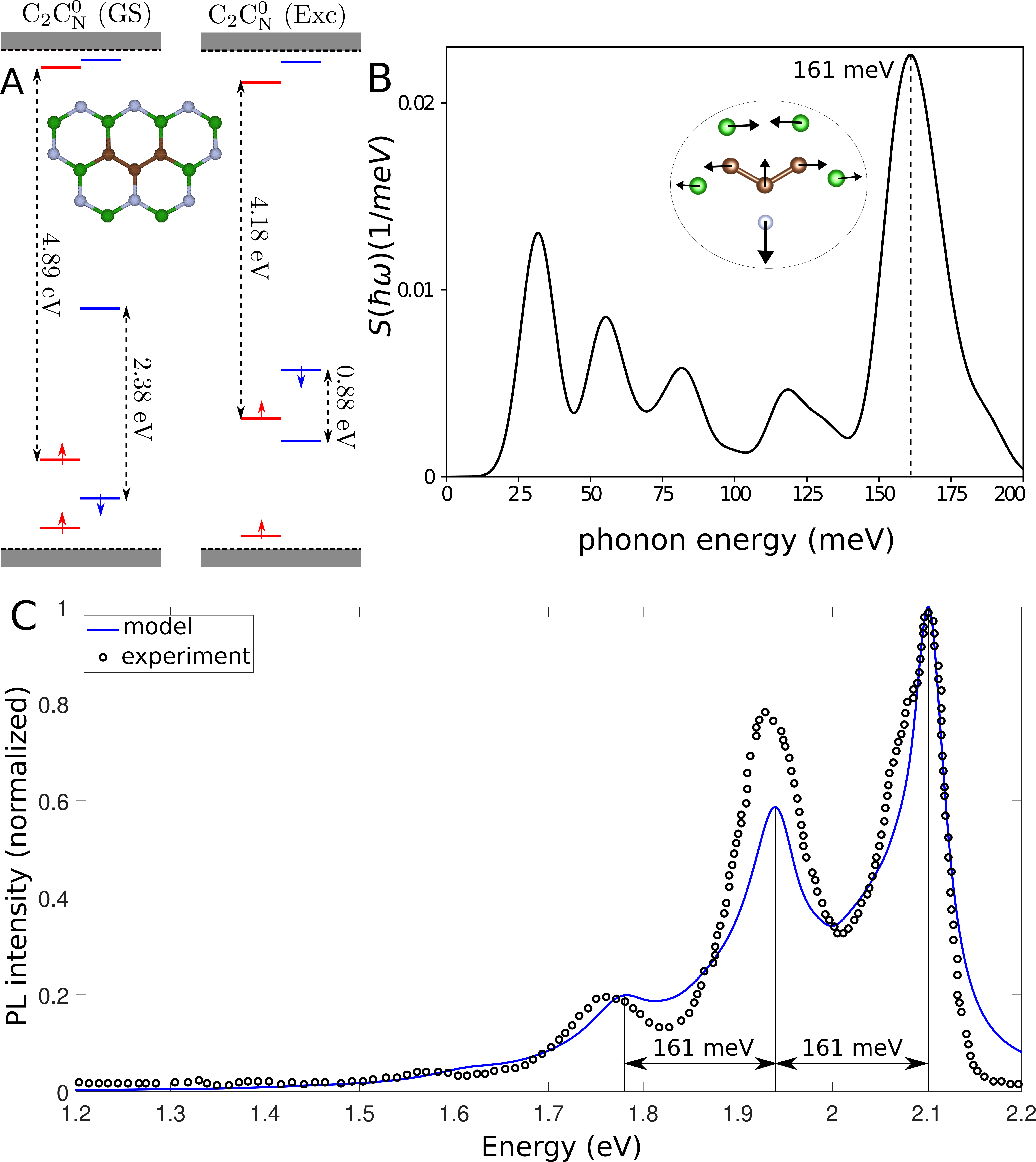}
    \caption{(a) Scheme of the defect levels of the neutral C$_2$C$_\mathrm{N}$ defect in the ground and excited level, the defect geometry is in the inset. (b) Spectral density $S(\omega)$ of this defect. The inset sketches the phonon at $\sim 160$ meV. (c) The calculated PL spectrum, together with a typical experimental spectrum (see Ref.~\cite{Martinez16}). The experimental ZPL was used.}
    \label{fig:trimers}
\end{figure}

The neutral trimers C$_2$C$_\mathrm{B}^0$ and C$_2$C$_\mathrm{N}^0$ are -qualitatively- similar to the C$_\mathrm{N}^0$ and C$_\mathrm{B}^0$ monomers, with a spin-split half-filled defect level which is closer to either the valence or conduction band. Unlike the monomers, there are extra defect states, similar to the case of C$_2^0$, see Fig.~\ref{fig:monomers}b. These additional states are very close to the conduction band minimum and valence band maximum. The trimer C$_2$C$_\mathrm{N}^0$ has a transition energy of $E_{ZPL}=1.62$ eV, calculated with HSE06. Its electronic structure is sketched in Fig.~\ref{fig:trimers}a for the ground and excited levels.  The accuracy of the calculation of the ZPL with DFT is indeterminated and should be taken as a hint to associate the defect with a SPE. Conversely, the spectral function $S(\omega)$, Fig.~\ref{fig:trimers}b, shows a single peak at $\sim$160 meV, which is the typical energy between the peaks of the PSB found in experiments \cite{Martinez16,tran16,Wigger19,kozawa2019}. For the sake of comparison between the PL spectrum of our model with the experiments \cite{Martinez16}, we shifted $E_{ZPL}$ to $2.1$ eV finding a good agreement with the PSB of this SPE. Therefore, the C$_2$C$_\mathrm{N}^0$ defect can be assigned as a SPE, likely with a ZPL close to 1.6 eV. Nevertheless, it is worth keeping in mind that the ZPL energy obtained with HSE06 is not an accurate value (\textit{i.e.} it can have an error of tenths of eV, and even be a match for the 2.1 eV SPE). However, we will use this value for the $E_{ZPL}$ in the lack of a better estimation. 

The other neutral trimer, C$_2$C$_\mathrm{B}^0$, has a similar electronic structure to C$_2$C$_\mathrm{N}^0$, with three defect levels within the fundamental band gap: one close to the valence band and fully occupied, one close to the conduction band and empty, and a third, half-occupied level, whose energy is in between the fully occupied and empty levels, see Fig.~\ref{fig:trimers}a. But, unlike the C$_2$C$_\mathrm{N}^0$ cluster, in C$_2$C$_\mathrm{B}^0$, the half-occupied level is shifted closer to the conduction band, \textit{i.e.} the energy between the last occupied and the first unoccupied levels are 2.19 and 5.03, for the spin majority and minority (red and blue in the figure), respectively. 
This cluster has a ZPL energy of 1.65 eV, very similar to the defect C$_2$C$_\mathrm{N}^0$. Also, their spectral function $S(\omega)$ and PL spectrum are very similar to C$_2$C$_\mathrm{N}^0$ (see Fig.~\ref{fig:trimers}). Therefore, both neutral C trimers are likely to be identified as SPEs in the range we are interested.

\subsection{Carbon tetramers}

There are few possible geometries for C tetramer defects. Discarding the ones with $C_3$ symmetry, only two possibilities remain, see Fig.~\ref{fig:tri-tetra}. As they have the same number of C$_\mathrm{N}$ and C$_\mathrm{B}$ substitutions, the neutral defect C$_4^0$ has a large separation between the empty and occupied defect states, similar to C$_2^0$ (see Fig.~\ref{fig:monomers}b). In the charged clusters C$_4^+$, C$_4^-$ the first optical transition has a (vertical) energy of less than 1 eV. Therefore, the tetramers are unlikely the observed SPEs.

\section{Double defects}
\label{sec:doublecluster}

\begin{figure}
    \centering
    \includegraphics[width=0.8\columnwidth]{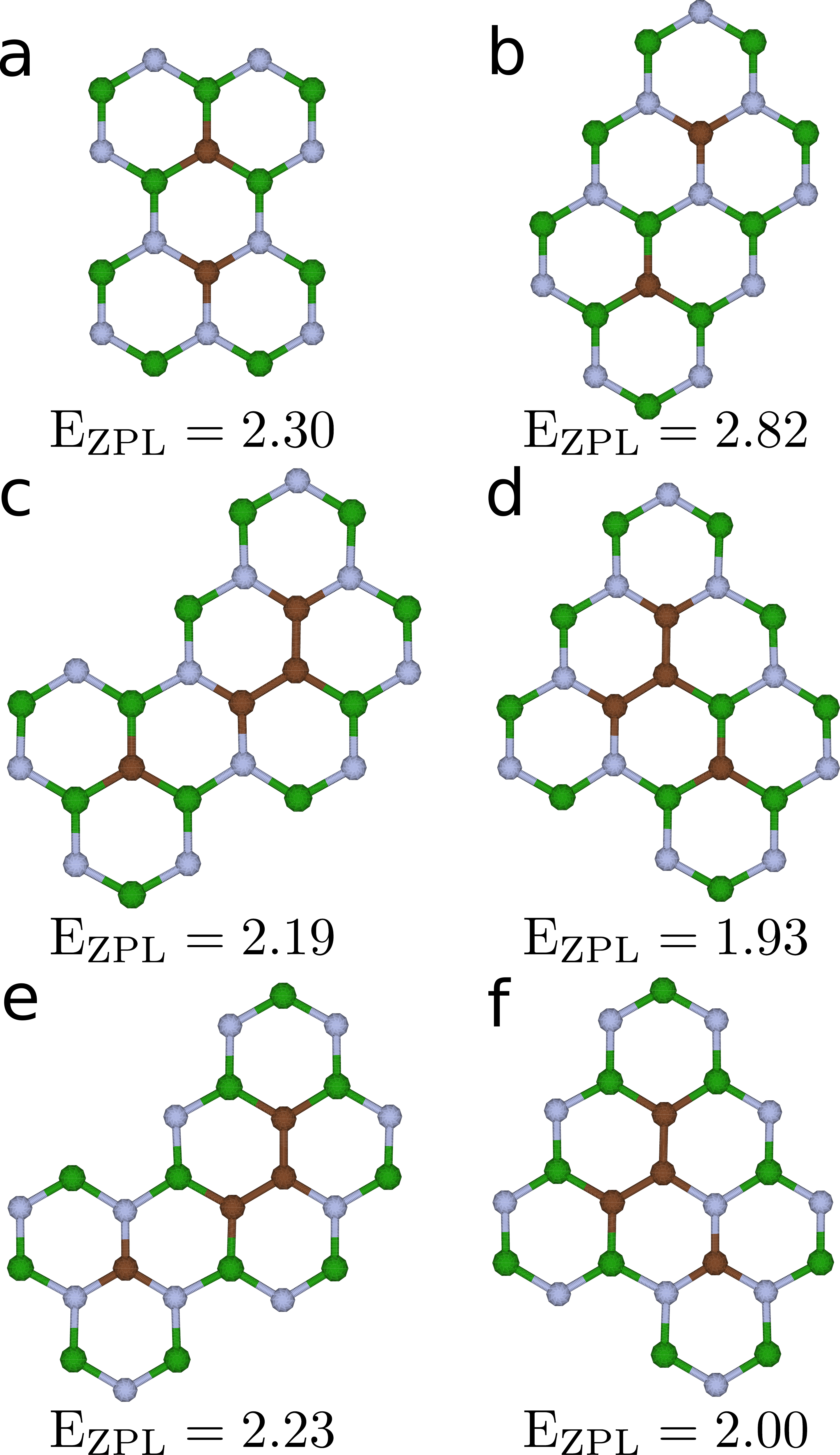}
    \caption{Different conformations of double defects C$_\mathrm{N}^--$C$_\mathrm{B}^+$ (a,b), C$_2$C$_\mathrm{B}^+-$C$_\mathrm{N}^-$ (c,d) and C$_2$C$_\mathrm{N}^--$C$_\mathrm{B}^+$ (e,f). The energy E$_{\mathrm{ZPL}}$ is in eV.} 
    \label{fig:double}
\end{figure}

From the previous discussion on C monomers and dimers, we can infer that two oppositely charged monomers (C$_\mathrm{B}^+$, C$_\mathrm{N}^-$) separated by some distance should have a similar set of defect levels to the $C_2^0$ dimer (Fig.~\ref{fig:monomers}b), but with a smaller separation between the occupied and empty levels (\textit{i.e.}, a weaker bonding). Fig.~\ref{fig:double}a,b shows two of such arrangements. The standard notation for defects is unable to distinguish between different arrangements. In what follows, we will label double defects according to Fig.~\ref{fig:double}. In this case, the C$_\mathrm{N}^--$C$_\mathrm{B}^+$ clusters will be named C$_\mathrm{N}^--$C$_\mathrm{B}^+-$a and C$_\mathrm{N}^--$C$_\mathrm{B}^+-$b. 

\begin{figure}
    \centering
    \includegraphics[width=0.92\columnwidth]{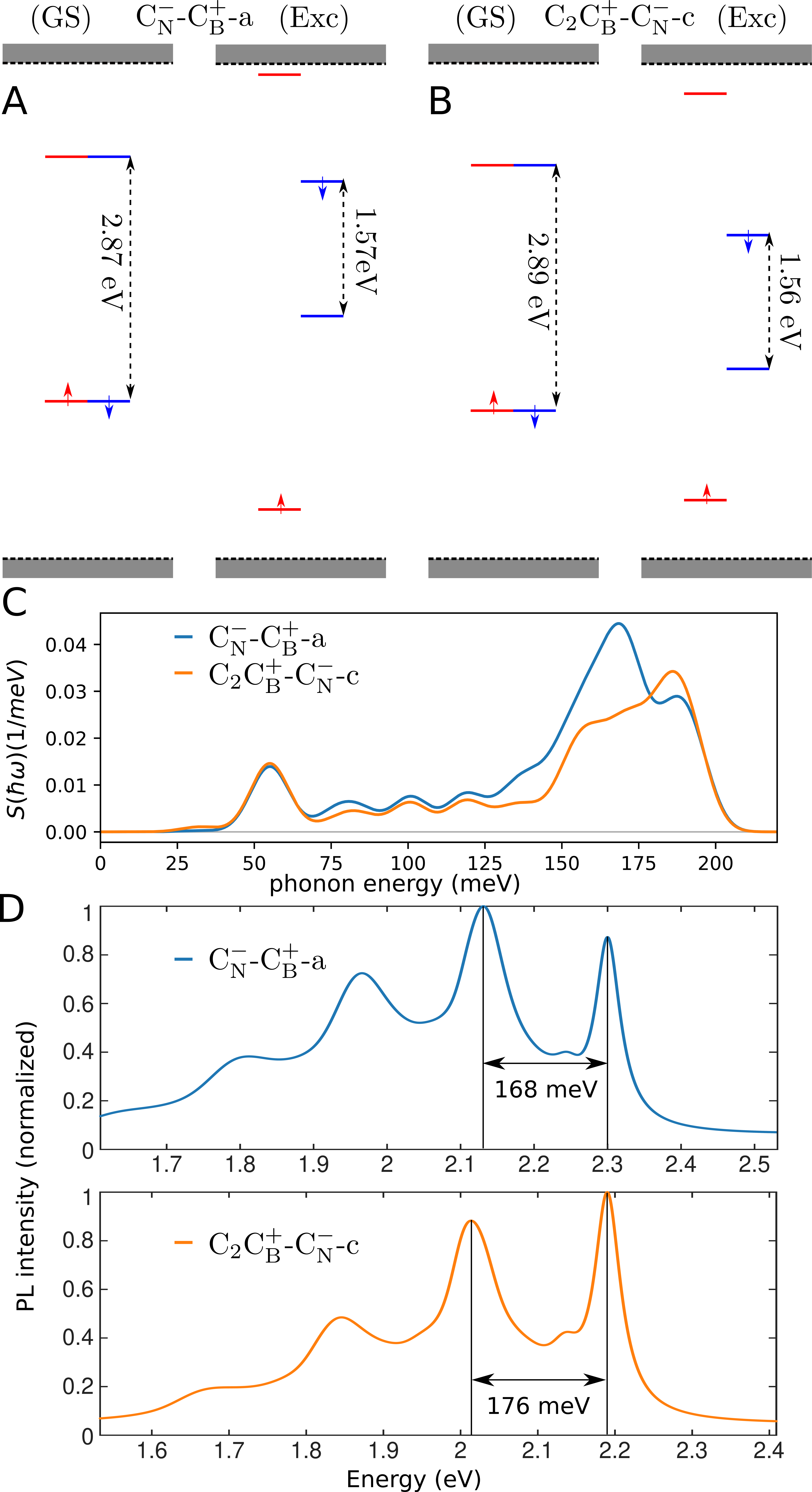}
    \caption{Scheme of the relevant defect levels of C$_\mathrm{N}^--$C$_\mathrm{B}^+-$a (panel a), and C$_2$C$_\mathrm{B}^+-$C$_\mathrm{N}^-$-c (panel b). Their respective spectral functions are in panel b and the calculated PL spectra are in panel c.}
    \label{fig:double-elec}
\end{figure}

The energy between the occupied and unoccupied defect states in the band gap of C$_\mathrm{N}^--$C$_\mathrm{B}^+$ quickly increases with the distance between the monomers. Only the cluster C$_\mathrm{N}^--$C$_\mathrm{B}^+-$a has a ZPL energy compatible with the observed spectrum ($E_{\mathrm{ZPL}}=2.30$ eV). The energy levels of this double cluster are sketched in Fig~\ref{fig:double-elec}a. Its corresponding spectral function $S(\omega)$ has a maximum at $\sim 170$ meV, in agreement with the observed PL spectra.

A very similar electronic structure is obtained by considering a charged trimer close to an oppositely charged monomer. The possible conformations keeping the trimer and the monomer separated by one or two atomic sites are shown in Fig.~\ref{fig:double}c-f. In all such conformations, the ZPL energy is compatible with the observed PL spectra. The electronic transitions of one case, C$_2$C$_{\mathrm{B}}^+$-C$_{\mathrm{N}}^{-}-$c, are shown in Fig.~\ref{fig:double-elec}b. The remaining cases are very similar. The spectral function of this cluster has a maximum at $~\sim 175$ meV. This combination of ZPL and phonon shift is compatible with some PL spectra \cite{Wigger19}. A local maximum is observed at $\sim 50$ meV. The prominence of these low-energy phonons also are in agreement with the measured PL. The remaining clusters of Fig.~\ref{fig:double} have a very similar spectral function and electronic structure (within the h-BN band gap). The corresponding peaks of the phonon shift in the PSB are: 172, 176 and 170 meV for C$_2$C$_{\mathrm{B}}^+$-C$_{\mathrm{N}}^{-}-$d, C$_2$C$_{\mathrm{N}}^-$-C$_{\mathrm{B}}^{+}-$e and C$_2$C$_{\mathrm{N}}^-$-C$_{\mathrm{B}}^{+}-$f, respectively.

\section*{Conclusions}

Using density functional theory calculations, we simulated the PL spectrum of several C defects in h-BN, up to C tetramers. Although, the zero-phonon line energy was calculated with the hybrid functional HSE06 (the \textit{de facto} standard for electronic structure calculations of solids), special attention was given to reproduce the prominent shift of the phonon sideband ($\sim 160$ meV) observed in most measurements of single-photon emitters in h-BN. This phonon sideband shift can only be originated by high-energy phonons, such as those involving bond stretching. Experiments also show (\textit{i}) a strong stark effect that cannot be attributed to defects with $C_3$ symmetry, and (\textit{ii}) that carbon impurities are needed to observe single-photon emitters in h-BN.

We found that the presence of vacancies is incompatible with most observed PL spectra\cite{tran16,Wigger19,kozawa2019,Martinez16,Du2015,Mendelson2020,Jung17}. After an optical transition, the most important phonons of a vacancy correspond to breathing modes, with an energy of $\sim 50$ meV. This behavior should also be present in more complex defects involving one or more vacancies.

With all the previous constraints, the natural options for single-photon emitters are carbon clusters formed by substitutional defects. We found that the neutral C trimers ($[\mathrm{C}_N]_2\mathrm{C}_B$, $\mathrm{C}_N[\mathrm{C}_B]_2$) have a phonon sideband similar to the one observed in typical PL experiments. The zero-phonon line of these clusters was calculated as $1.63$ and $1.65$ eV, respectively.

We also studied the case of two C clusters next to each other or double clusters. The most interesting case is formed by two oppositely charged C clusters (\textit{e.g.} C$_\mathrm{B}$--C$_\mathrm{N}$). The overall charge state of these double clusters is neutral, similarly to the case of the C dimer (responsible for the 4.1 eV single-photon emitter), but with a lower zero-phonon line energy. We found five atomic arrangements of double clusters with a zero-phonon line within the visible range. In all these cases, the phonon sideband shows a strong shift of $\sim 160$, in agreement with experiments.

The existence of at least two different types of singe-photon emitters in h-BN, with very different underlying physics, has been observed \cite{hayee2020}. Our findings give a coherent explanation for this behavior. Nevertheless, the existence of many defects optically active in h-BN may hinder its potential for applications and further research is needed to learn how to select a single type of SPEs.

\begin{acknowledgments}
This work was partially supported by Fondecyt Grants No. 1191353 and 1180673, by the  
Center for the Development of Nanoscience and
Nanotechnology CEDENNA AFB180001 and from Conicyt PIA/Anillo ACT192023. 
This research was partially supported by the supercomputing infrastructure 
of the NLHPC (ECM-02). AN acknowledges financial support from Universidad Mayor through the Postdoctoral Fellowship. TR and SB acknowledge financial support from the Deutsche Forschungsgemeinschaft (DFG, German Research Foundation)
through the projects SFB-1375 (project A02) and BO 4280/8-1 and 
from the Volkswagen Stiftung (Momentum) through the project ``dandelion''. 
\end{acknowledgments}




\bibliography{bib}

\end{document}